\documentclass[useAMS,usenatbib,usegraphicx]{mn2e}
\usepackage{amsmath}
\usepackage{pgf}
\bibpunct{(}{)}{;}{a}{,}{,}

\input epsf

\makeatletter
\def\ref@jnl#1{{\rmfamily#1}}%
\newcommand\aj{\ref@jnl{AJ}}%
\newcommand\araa{\ref@jnl{ARA\&A}}%
\newcommand\apj{\ref@jnl{ApJ}}%
\newcommand\apjl{\ref@jnl{ApJ}}%
\newcommand\apjs{\ref@jnl{ApJS}}%
\newcommand\ao{\ref@jnl{Appl.~Opt.}}%
\newcommand\apss{\ref@jnl{Ap\&SS}}%
\newcommand\aap{\ref@jnl{A\&A}}%
\newcommand\aapr{\ref@jnl{A\&A~Rev.}}%
\newcommand\aaps{\ref@jnl{A\&AS}}%
\newcommand\azh{\ref@jnl{AZh}}%
\newcommand\baas{\ref@jnl{BAAS}}%
\newcommand\jrasc{\ref@jnl{JRASC}}%
\newcommand\memras{\ref@jnl{MmRAS}}%
\newcommand\mnras{\ref@jnl{MNRAS}}%
\newcommand\pra{\ref@jnl{Phys.~Rev.~A}}%
\newcommand\prb{\ref@jnl{Phys.~Rev.~B}}%
\newcommand\prc{\ref@jnl{Phys.~Rev.~C}}%
\newcommand\prd{\ref@jnl{Phys.~Rev.~D}}%
\newcommand\pre{\ref@jnl{Phys.~Rev.~E}}%
\newcommand\prl{\ref@jnl{Phys.~Rev.~Lett.}}%
\newcommand\pasp{\ref@jnl{PASP}}%
\newcommand\pasj{\ref@jnl{PASJ}}%
\newcommand\qjras{\ref@jnl{QJRAS}}%
\newcommand\skytel{\ref@jnl{S\&T}}%
\newcommand\solphys{\ref@jnl{Sol.~Phys.}}%
\newcommand\sovast{\ref@jnl{Soviet~Ast.}}%
\newcommand\ssr{\ref@jnl{Space~Sci.~Rev.}}%
\newcommand\zap{\ref@jnl{ZAp}}%
\newcommand\nat{\ref@jnl{Nature}}%
\newcommand\iaucirc{\ref@jnl{IAU~Circ.}}%
\newcommand\aplett{\ref@jnl{Astrophys.~Lett.}}%
\newcommand\apspr{\ref@jnl{Astrophys.~Space~Phys.~Res.}}%
\newcommand\bain{\ref@jnl{Bull.~Astron.~Inst.~Netherlands}}%
\newcommand\fcp{\ref@jnl{Fund.~Cosmic~Phys.}}%
\newcommand\gca{\ref@jnl{Geochim.~Cosmochim.~Acta}}%
\newcommand\grl{\ref@jnl{Geophys.~Res.~Lett.}}%
\newcommand\jcp{\ref@jnl{J.~Chem.~Phys.}}%
\newcommand\jgr{\ref@jnl{J.~Geophys.~Res.}}%
\newcommand\jqsrt{\ref@jnl{J.~Quant.~Spec.~Radiat.~Transf.}}%
\newcommand\memsai{\ref@jnl{Mem.~Soc.~Astron.~Italiana}}%
\newcommand\nphysa{\ref@jnl{Nucl.~Phys.~A}}%
\newcommand\physrep{\ref@jnl{Phys.~Rep.}}%
\newcommand\physscr{\ref@jnl{Phys.~Scr}}%
\newcommand\planss{\ref@jnl{Planet.~Space~Sci.}}%
\newcommand\procspie{\ref@jnl{Proc.~SPIE}}%

\makeatother

\def\vecw{{\bf w}}

\long\def\comment#1{}

\def\W2{{\cal W}}

\def\dalemb#1#2{{\vbox{\hrule height.#2pt
        \hbox{\vrule width.#2pt height#1pt \kern#1pt \vrule width.#2pt}
        \hrule height.#2pt}}}

\def\ba{\begin{eqnarray}}
\def\ea{\end{eqnarray}}
\def\be{\begin{equation}}
\def\ee{\end{equation}}
\def\bda{\boldsymbol{a }}
\def\bdn{\boldsymbol{n }}
\def\bds{\boldsymbol{s }}
\def\bdw{\boldsymbol{w }}
\def\bdx{\boldsymbol{x }}
\def\bdelta{\boldsymbol{\delta }}

\def\gtorder{\mathrel{\raise.3ex\hbox{$>$}\mkern-14mu
             \lower0.6ex\hbox{$\sim$}}}
\def\ltorder{\mathrel{\raise.3ex\hbox{$<$}\mkern-14mu
             \lower0.6ex\hbox{$\sim$}}}

\newcommand{\tR}{{\rm R}}
\newcommand{\tA}{{\rm A}}

\newcommand{\tO}{{\rm O}}
\newcommand{\tD}{{\rm D}}

\newcommand{\var}{{\rm var}}

\def\be{\begin{equation}}
\def\ee{\end{equation}}
\def\bea{\begin{eqnarray}}
\def\eea{\end{eqnarray}}

\def\cmm2{{\,\rm cm^{-2}}}
\def\cm2{{\,{\rm cm}^2}}
\def\cmm3{{\,{\rm cm}^{-3}}}
\def\gcmm3{{\,{\rm g\,cm^{-3}}}}

\def\fun#1#2{\lower3.6pt\vbox{\baselineskip0pt\lineskip.9pt
  \ialign{$\mathsurround=0pt#1\hfil##\hfil$\crcr#2\crcr\sim\crcr}}}

\title[Calibration Errors and CMB Component Separation]{Impact of calibration errors on CMB component separation using FastICA
  and ILC}
\author[Jason Dick, Mathieu Remazeilles, Jacques Delabrouille]{Jason Dick$^1$\thanks{E-mail: dick@sissa.it}, Mathieu Remazeilles$^{2,3}$\thanks{E-mail: remazeil@apc.univ-paris7.fr}, Jacques Delabrouille$^2$\thanks{E-mail: delabrouille@apc.univ-paris7.fr}\\
$^{1}$ SISSA
via Beirut, 2-4,
34014 Trieste,
Italy\\
$^2$ APC
10, rue Alice Domon et L\'eonie Duquet,
75205 Paris Cedex 13,
France\\
$^3$ Laboratoire de Physique Th\'eorique, Universit\'e Paris-Sud, 
91405 Orsay, 
France}

\begin{document}

\date{Submitted to MNRAS 2009 July 17}

\pagerange{\pageref{firstpage}--\pageref{lastpage}} \pubyear{2009}

\maketitle

\label{firstpage}

\begin{abstract}
  The separation of emissions from different astrophysical processes is an
  important step towards the understanding of observational data.  This topic
  of component separation is of particular importance in the observation of
  the relic Cosmic Microwave Background Radiation, as performed by the WMAP
  satellite and the more recent Planck mission, launched May 14th, 2009 from
  Kourou and currently taking data.  When performing any sort of component
  separation, some assumptions about the components must be used.  One
  assumption that many techniques typically use is knowledge of the frequency
  scaling of one or more components.  This assumption may be broken in the
  presence of calibration errors.  Here we compare, in the context of
  imperfect calibration, the recovery of a clean map of emission of the Cosmic
  Microwave Background from observational data with two methods: FastICA
  (which makes no assumption of the frequency scaling of the components), and
  an `Internal Linear Combination' (ILC), which explicitly extracts a
  component with a given frequency scaling.  We find that even in the presence
  of small calibration errors with a Planck-style mission, the ILC method can
  lead to inaccurate CMB reconstruction in the high signal-to-noise regime,
  because of partial cancellation of the CMB emission in the recovered map.
  While there is no indication that the failure of the ILC will translate to
  other foreground cleaning or component separation techniques, we propose
  that all methods which assume knowledge of the frequency scaling of one or
  more components be careful to estimate the effects of calibration errors.
\end{abstract}


\section{Introduction}

Precise observation of the Cosmic Microwave Background (CMB), as
planned with the Planck space mission
\citep{bersanelli00, lamarre00, tauber04}, is
of the utmost importance for better understanding, and confronting
with precise observational data, the hot big bang model and its
theoretical predictions. In this theoretical framework, such
observations also permit constraining the parameters of the model, as
is currently done to a lesser extent by a number of previous
experiments, such as COBE \citep{fixsen97}, WMAP \citep{komatsu09},
ACBAR \citep{reichardt09}, Archeops \citep{beno03,tristram05}, BOOMERANG
\citep{mactavish06}, CBI \citep{sievers09}, QUaD \citep{quad08}, and VSA
\citep{rebolo04}.

With ever more sensitive instruments, the main source of uncertainty
in CMB observations, rather than being instrumental noise, is the
contamination of the observation by foreground emission.
Astrophysical foregrounds comprise millimeter wave emission from the
interstellar medium in our own galaxy, as well as emission from
compact extragalactic sources.

Component separation methods make use of the different emission laws of
different astrophysical components to separate them through joint analysis of
observations made at different wavelengths \citep{delabrouille07}. Among those
methods, the so-called Internal Linear Combination (ILC), which makes few
assumptions about the physical properties of the CMB and the foregrounds, has
been widely used for the analysis of WMAP data
\citep{tegmark03,eriksen04,delabrouille09,kim09}.  An important assumption of
the ILC is that the frequency scaling of the CMB is assumed to be known. This
is, in principle, a safe assumption, as small temperature fluctuations $\Delta
T$ of the CMB generate brightness fluctuations proportional to $\Delta T$,
which scale in frequency like the derivative of a blackbody with respect to
the temperature, at the well measured CMB temperature of $T=2.725$ K. However,
calibration coefficients for each channel, which are a multiplicative factor
for each frequency, introduce an uncertainty in the frequency scalings of the
CMB component in presence of calibration errors. For space-based missions
these uncertainties are typically small (well below 1\% for WMAP or Planck).

More sophisticated methods for component separation have been extensively
studied in the community of statistical signal processing for a variety of
applications. These methods are part of a field of activity generically
designated as Blind Source Separation (BSS), or equivalently Independent
Component Analysis (ICA). ICA methods perform separation on the basis of the
assumption that each of the available observations is a different linear
mixture of a well defined number of statistically independent components. Such
methods generically rely on no prior assumption on the scaling coefficients of
the components in the different available observations (i.e. on the
coefficients of each component in the `mixtures').  In fact, recovering these
coefficients (the so--called `mixing matrix') is precisely the primary target
of blind source separation.  ICA methods, thus, do not typically assume
perfect knowledge of the response of each channel to the CMB -- nor that the
CMB contribution is the same in all channels. For CMB studies particularly,
this type of approach has led to the development of a large variety of
methods, including CCA \citep{bonaldi06, bedini05}, FastICA
\citep{hyvarinen99}, SMICA \citep{delabrouille03, cardoso08} and GMCA
\citep{bobin08}. These methods have been used on real observational data in a
variety of contexts \citep{bonaldi07, maino06, patanchon05}, and compared
extensively on simulated data sets \citep{leach08}.

The two main differences between the ILC and ICA methods are the following:
\begin{itemize}
\item Whereas ICA is designed to extract the scaling coefficients of
  each of the identified components from the data themselves, the ILC
  assumes perfect knowledge of the scaling coefficients for the
  component of interest (CMB);
\item The ILC does not make any assumption about the properties of
  foreground contamination, whereas ICA assumes that the data are
  satisfactorily described by a (noisy) linear mixture of independent
  components.
\end{itemize}

Clearly, these methods are bound to be more or less adapted to
component separation, depending upon the actual properties of the data
set and on the science objectives pursued. In the following we
propose to investigate, using realistic simulations of sky emission
and of observational data for WMAP and Planck, the
relative performance of FastICA and ILC in the presence of calibration
errors. Such calibration errors result in the violation of one of the
assumptions of the ILC (the prior knowledge of the exact scaling
coefficients of the CMB in the observations). By contrast, blind
component separation methods are designed from first principles to
estimate the scaling coefficients from the data, and in principle
should not suffer much from calibration uncertainties.

The rest of this paper is organized as follows. In section
\ref{sec:ilcica} we describe the ILC and ICA component separation methods. 
We describe our methodology for comparing
the methods in section \ref{sec:method}.  In section
\ref{sec:results} we present the results of our analysis, followed
by our conclusions in section \ref{sec:conclusion}. We also provide
a detailed calculation of the effect of calibration errors on the
ILC in the appendix \ref{app:ilcweights}.

\section{ILC and ICA}\label{sec:ilcica}

In the following we assume that the available data (maps $x_i(p)$ of
observed sky) can be written as
\begin{equation}
x_i(p) = a_i s(p) + n_i(p),
\end{equation}
where $s(p)$ is the map of the component of interest (the CMB), $p$
indexes pixels in the map, and $n_i(p)$ is the contribution from
foregrounds and instrumental noise to the map $x_i(p)$. The
coefficients $a_i$ scale the relative amplitude of the CMB map in the
different available observations. For observations in thermodynamic
units, and perfect calibration, we have $\forall i$, $a_i = 1$.

\subsection{The ILC}
The philosophy behind the ILC is to find the linear combination of the
available maps $x_i$ which has minimal variance while retaining unit
response to the CMB map. This linear combination, $\sum_i w_i x_i(p)$,
is then an estimate $\hat s(p)$ of the true CMB map $s(p)$.  The ILC
weights $w_i$ are found by solving the problem of minimizing
$\var{\sum_i w_i x_i(p)}$ under the constraint $\sum_i w_i = 1$. In
principle, this last constraint guarantees unit response to the CMB,
as we have:

\begin{eqnarray}
\hat s(p) &=& \sum_i w_i x_i(p) \cr
&=& s(p) + \sum_i w_i n_i(p).
\end{eqnarray}

In the presence of foregrounds, which induce correlated errors from
channel to channel, the ILC weights adjust themselves so that the
linear combination cancels out as much of the foregrounds as
possible. The actual weights, however, result from a trade-off between
canceling foregrounds and allowing errors due to instrumental noise in
the final map.

The constrained minimization problem can be solved in a straightforward
manner using a Lagrange multiplier method to impose $\sum_i w_i =
1$. The resulting weights are found to be:
\begin{equation}
\vecw = \frac{{\widehat{\tR}}^{-1} \, \bda}{\bda^t \,
  {\widehat{\tR}}^{-1} \, \bda},
\end{equation}
where ${\widehat{\tR}}$ is the empirical covariance matrix of the
observations.  Note that we have used bold font to denote vectors, and
have omitted the reference to the pixel value.  From here on, this
notation will be used.  The ILC estimator of the CMB map $s(p)$ can be
written as:

\begin{equation}
\hat{s}_{\rm ILC} = \vecw^t \bdx =  \frac{\bda^t \, {\widehat{\tR}}^{-1}}{\bda^t \,
  {\widehat{\tR}}^{-1} \, \bda} \, \bdx.
\label{eq:ILC}
\end{equation}

The ILC weights, obviously, depend upon the assumed scaling
coefficients $a_i$ for the component of interest. It is then clear
that an error in the assumed scalings changes the ILC performance, but
by how much?  As the ILC attempts to minimize the total variance of
the output map, the constraint that $\sum w_i a_i = 1$ plays a
critical role in guaranteeing that the linear combination does not
adjust its coefficients to cancel the CMB as well as foregrounds. It
is foreseeable, then, that calibration errors could, in some cases,
impact the performance of ILC more severely than just a small overall
calibration error on the final output map.

\subsection{FastICA}

There is a wide choice of possible ICA methods to extract the CMB from
multifrequency observations. In this paper, we make use of the
standard FastICA algorithm as described in \citet{hyvarinen99}, with a
few minor changes: 
\begin{itemize}
\item We subtract an estimate of the instrument noise covariance matrix from
  the empirical covariance matrix of the data.
\item Instead of leaving the estimated signal as being unit variance, we set
  the CMB scaling to be such that the sum of the weights is equal to one,
  mirroring the ILC method to ensure unit response to the CMB.
\end{itemize}

FastICA is based on the general principle that a sum of two different
independent probability distributions will always tend to be more
Gaussian than either of the distributions are independently.  We can
thus extract $N$ independent sources from $N$ channels of data by
forming the linear combination of the $N$ channels which maximizes the
non-Gaussianity of the extracted sources.  A measure of the
non-Gaussianity of each source is performed using functions such as:

\begin{equation}
Y(x) \propto \left[E\left\{G(x)\right\} - E\left\{G(y)\right\}\right]^2.
\end{equation}

where $x$ is data that has unit variance, and $y$ is a random variable drawn
from a unit-variance Gaussian distribution.  Here $E\{\}$ is the expectation
value of the data set or probability distribution enclosed and $G(x)$ is some
non-linear function.  Popular choices include a Gaussian, a polynomial, or the
logarithm of the hyperbolic cosine.  Which specific choice is best depends
upon precisely how the distribution of $x$ differs from a Gaussian, though it
is clear that for any choice of $G(x)$, $Y(x)$ will be zero if $x$ is
Gaussian-distributed, and positive definite otherwise.  In the present paper,
we use $G(x) = x^4$.

FastICA assumes a model of the data of the form :
\begin{equation}
\bdx = \tA\bds + \bdn,
\label{eqn:datamodel}
\end{equation}
where now vector $\bds$ comprises all `sources' (CMB + foregrounds), and
$\bdn$ is instrumental noise only (for all channels).  The objective of the
method is to evaluate the mixing matrix $\tA$, and then use this estimate to
invert the linear system.

In order to optimize estimation of the mixing matrix that determines
the linear combination of $x$ which represents the individual sources,
FastICA also performs a pre-whitening step.  This pre-whitening step
exploits the assumption of statistical independence to perform a
linear transformation on the data, which sets its covariance matrix to
the identity by multiplying the data by the inverse square root of its
covariance.  The mixing matrix then becomes a simple rotation matrix
which, with its smaller number of degrees of freedom, is easier to
estimate.

For generating the pre-whitening matrix, we do not make direct use of the
covariance matrix of the data, as with basic FastICA, but instead use the
estimated covariance matrix of the signal as in \citet{maino02}.  This can be
understood simply by our modeling of the data (equation \ref{eqn:datamodel}).
Given this data model, the covariance of the observations is:
\begin{eqnarray}
\tR_x &=& \left<\left(\tA\bds + \bdn\right)\left(\tA\bds + \bdn\right)^t\right>\cr
\tR_x &=& \tA\tR_s\tA^t + \tR_n.
\end{eqnarray}

Here the correct covariance matrix to use to whiten the signal is
$\tA\tR_s\tA^t$, which we estimate as $\tR_x - \tR_n$.  The channel-channel
noise covariance $\tR_n$ is taken as diagonal with the diagonal elements
estimated from our knowledge of the per-pixel noise in each map
combined with how much each map was smoothed.  We have assumed that
the signal and noise are uncorrelated in the above derivation.

Having performed the pre-whitening, all extracted sources have unit variance
and are uncorrelated.  To determine the overall CMB scaling, we first
determine which of the sources is the CMB, then use the ILC strategy of
setting the sum of the CMB weights equal to one.  This ensures that the level
of the CMB in the output is, at least in the case of no calibration error,
equal to the level of the CMB in the maps.

\section{Method}\label{sec:method}

We now turn to the investigation of the impact of calibration errors on
component separation with ILC and FastICA. The approach of this investigation
consists of generating simulated `observations', with varying calibration
errors, noise levels, and frequency channels, and compare the performance of
ILC and FastICA at recovering the CMB map.

Performance is measured in several ways, based on the measurement of
reconstruction errors of different types.

Denoting as $s(p)$ the (beam-smoothed) CMB map used in the simulation, and as
$\hat{s}(p)$ the CMB map obtained from processing the simulated data, the
reconstruction error is $\hat{s}(p) - s(p)$.

This reconstruction error arises from two terms. A multiplicative term (i.e. a
global calibration error) and an additive term.  We have
$$
\hat{s}(p) = \alpha s(p) + c(p)
$$ where $\alpha$ is the global calibration coefficient, and $c(p)$ the
additive contamination by foregrounds and noise.  Ideally, we aim at $\alpha =
1$ and $c(p) = 0$.

In practice, in both ILC and ICA methods, the final map is reconstructed as a
linear combination $\sum w_i x_i(p)$ of the input maps $x_i(p)$.  Hence, for
simulated data, one can compute easily $\alpha = \sum w_i a_i$ and $c(p) =
\sum w_i n_i(p)$, where $n_i(p)$ are maps of the sum noise and foregrounds in
channel $i$.

The comparison of the variance of the reconstruction error, of the overall
response $\alpha$, and of the contamination $c(p)$ for ILC and ICA gives
insight on the relative performance of the two, and of the main origin of
error, in presence of calibration uncertainties.

\subsection{Simulations}

In preparation for the forthcoming Planck space mission, simulations
for the 9 Planck frequency channels, from 30 to 857 GHz, as described
in the Planck 'Bluebook'\footnote{\tt \tiny
  {http://www.rssd.esa.int/SA/PLANCK/docs/Bluebook-ESA-SCI(2005)1\_V2.pdf}},
are made. We also consider simulations in the WMAP frequency channels,
between 23 and 94 GHz.  Sky simulations are performed using the Planck
Sky Model (PSM) package, version 1.6.3\footnote{\tt \tiny
  {http://www.apc.univ-paris7.fr/APC\_CS/Recherche/Adamis/PSM/psky-en.php}}
and using the Healpix pixelization. In the simulated observations, we
introduce a small calibration error, so that each of the sky maps is
multiplied by a calibration coefficient. We consider calibration
errors $\delta a/a$ of 0.1, 0.2, 0.5, and 1\%, which implies
calibration coefficients typically somewhere between 0.99 and
1.01.\footnote{The calibration error expected for Planck is less than
  1\% up to the 353GHz channel, as given by the Planck 'Bluebook'} We
work at the resolution of the lowest frequency channel in our
simulations, i.e. 33 arcminute beams for Planck, and 54 arcminute
beams for WMAP.

Noise compatible with what is expected for the two instruments, for
maps smoothed at the resolution of the lowest frequency channel, is
added to the sky emission.  We then separate components with both an
ILC and with FastICA, and analyze and interpret the results.

\subsubsection{Planck Sky Model}

Sky maps are generated using a four-component model of galactic emission which
includes free-free, synchrotron, thermal dust, and spinning dust diffuse
components.  We also add emission from several populations of compact sources,
which comprise ultracompact galactic H-II regions, infra-red and radio sources
(both galactic and extragalactic), a far infrared background emission, and
thermal SZ effect from a simulated distribution of galaxy clusters.  For our
Planck simulations, maps are generated at 30GHz, 44GHz, 70GHz, 100GHz, 143GHz,
217GHz, 353GHz, 545GHz, and 857GHz, each at nside=1024.  For WMAP simulations,
maps are generated at 23GHz, 33GHz, 41GHz, 61GHz, and 94GHz, each at
nside=512.  Maps are simulated using Gaussian symmetric beams.  Only
temperature maps are generated.

\subsubsection{Post-processing of PSM Outputs}
Instrumental noise is added separately after the sky is simulated with the
PSM.  For Planck, we assume uniform sky coverage, with noise level
corresponding to what is given in the Planck `Bluebook'.  Since the FastICA
and ILC methods require maps that are at the same resolution, we then smooth
all maps to the resolution of the 30GHz channel, which has a Gaussian beam
FWHM of 33'.  As we use a relatively low resolution beam, all maps are set to
nside=512 after smoothing.

After adding noise and smoothing maps to the same resolution, we simulate the
calibration error by drawing a zero-mean Gaussian random variable $x$ with RMS
equal to the desired calibration error (e.g. $\sigma=0.002$ for $0.2\%$
error).  We then multiply the map by $1 + x$.  This is repeated for each
frequency channel, with the same calibration RMS error but a different
realization of $x$ for each.

While it makes no difference whether the calibration error simulation is
performed before or after smoothing, we note that it is correct to add the
calibration error after the noise, as the overall estimated noise level also
depends upon the calibration of the instrument.  As we make use of the
estimated noise covariance between the channels, the estimated noise level
after smoothing is also computed here.

\subsection{Masking}

For better performance of the FastICA or ILC component separation
algorithms, it is safer to mask out particularly bright sources as
well as those with strongly-varying spectral properties.  The mask is
determined making use of a simple magnitude-based algorithm.  First,
we produce a theoretical estimate of the expected CMB RMS based upon
the WMAP power spectrum.  We then generate a mask that removes all
pixels which contain a value larger than four times the CMB RMS.

\begin{figure}
\includegraphics[width=85mm]{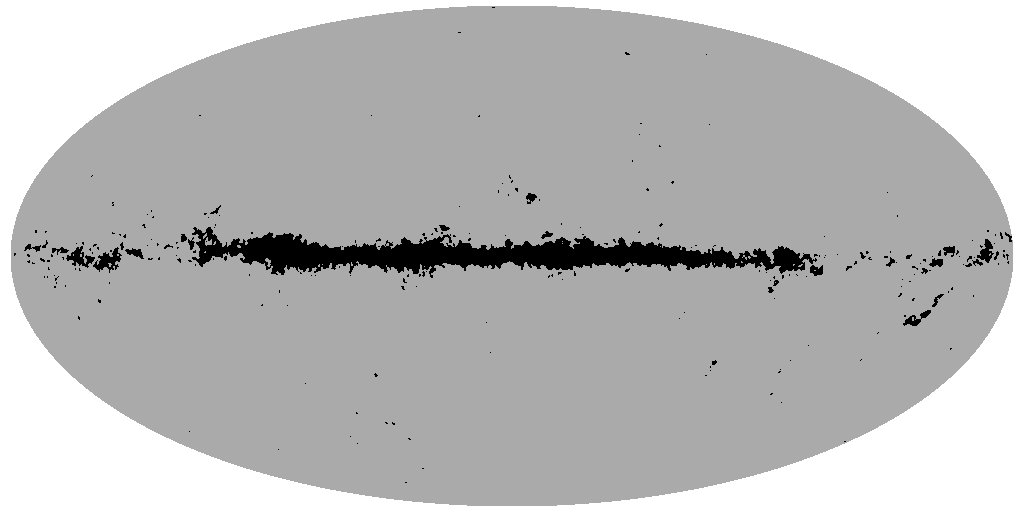}
\caption{Mask that removes the brightest pixels from the 70GHz and
  100GHz channels.}
\label{fig:mask}
\end{figure}

For our maps, the mask used is a union of the masks computed as above from the
70GHz and 100GHz channels.  We make use of the mask as generated from the
first realization with no calibration error, and do not recompute the mask
between runs.  The resultant mask is shown in fig. \ref{fig:mask}.  It is
possible that we could obtain better component separation performance through
more precise masking, but this is not expected to have any impact on the
overall results of the present paper. The study could have been performed with
any arbitrary mask, as long as the average CMB to foreground ratio is not
changed significantly.

\subsection{Monte Carlo}
In order to investigate both the average of the reconstruction error and its
dispersion, we individually execute each of the above steps many times for
each chosen set of parameters, the exact number depending upon the test.
Summary statistics are then computed across the runs.  When comparing
different component separation techniques, the exact same set of realizations
are used.  Different choices of the calibration error level also make use of
the same input sky maps.

For these simulations, CMB and noise are generated from their statistical
properties separately in each simulation.  The CMB is a Gaussian realization
assuming, for all simulations, the same power spectrum, compatible with WMAP
best fit model, but new phases for each realization. Similarly, all
realizations of noise are independent.

Other components are not fully independent from realization to
realization. Galactic components, the model of which is heavily
constrained by WMAP observations, do not change much. The Sunyaev
Zel'dovich map is fixed (i.e. the same SZ template map is used in all
simulations). A fraction of point sources remain similar (they are
based on the positions of real sources) although their spectral
emission law depends on the realization. An additional population of
point sources, generated to correct for the sky coverage of point
source surveys to homogenize the point source distribution, is
generated independently for each sky realization.

\section{Results}\label{sec:results}

In this section we present both analytical and numerical results obtained
after including the presence of calibration errors in the ILC and ICA
component separation methods. The success or the failure of a method will be
evaluated as follows. We construct the output CMB map estimates by ILC or ICA
as well as the residual map, which is the difference map between the estimated
output CMB map and the simulated input CMB map. We compute the RMS value of
each of these maps and compare them. We also evaluate both the multiplicative
factor $\alpha$ and the additive error $c(p)$ (introduced in section
\ref{sec:method}), characterizing the reconstruction errors.

\subsection{Compared reconstruction error}

\begin{figure}
\includegraphics[width=90mm]{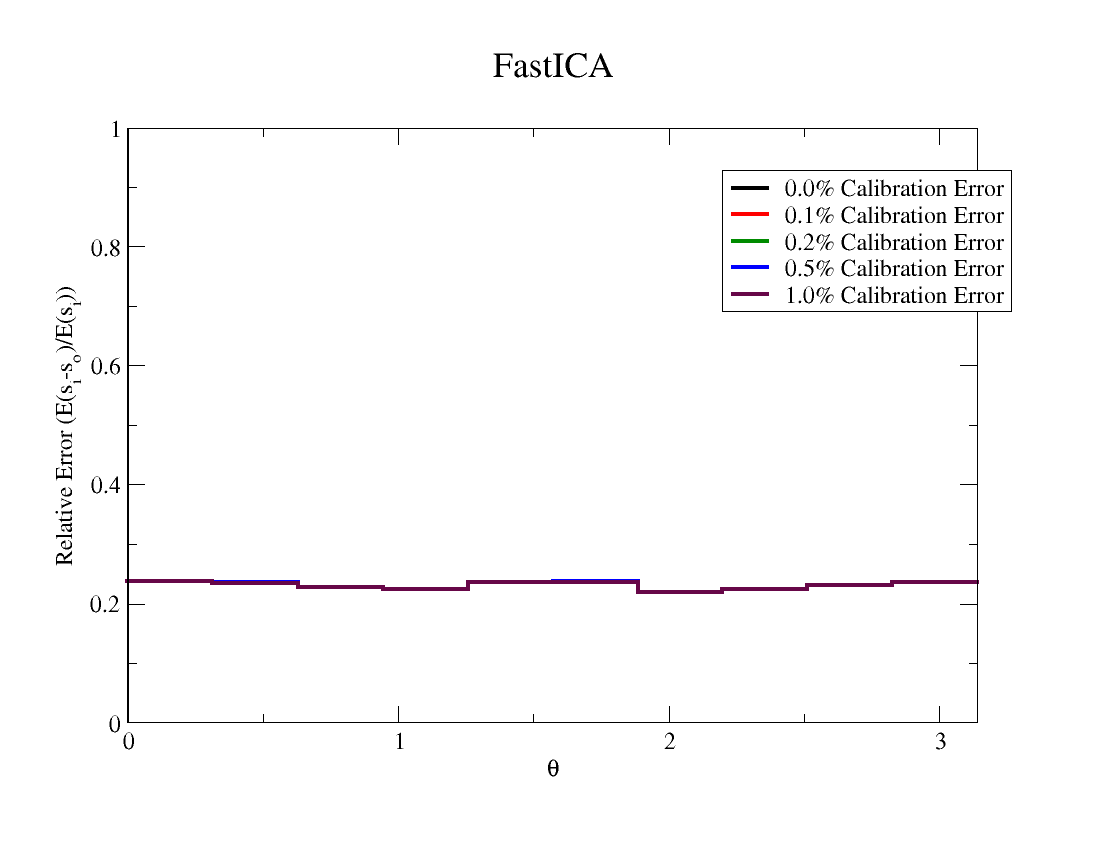}
\caption{Plot of the relative error of FastICA as a function of galactic
  latitude.  Generated using 128 simulations for each case.  As expected, the
  relative error of FastICA has very little dependence upon the calibration
  error.}
\label{fig:rel_err_ica}
\end{figure}

\begin{figure}
\includegraphics[width=90mm]{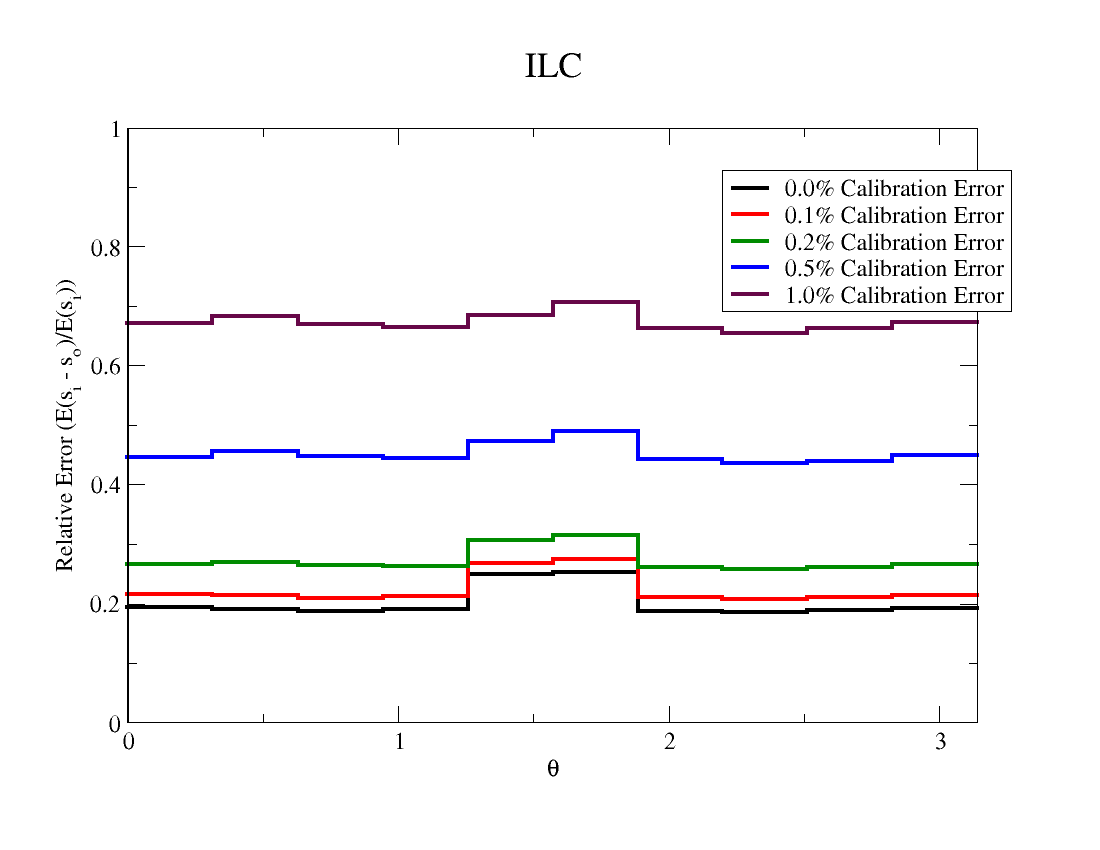}
\caption{Plot of the relative error of ILC as a function of galactic latitude.
  Generated using 128 simulations for each case.  Unlike FastICA, ILC shows
  tremendous sensitivity to the calibration error, causing a noticeable
  reduction in the quality of the extraction of the CMB even at the optimistic
  0.1\% calibration error level.}
\label{fig:rel_err_ilc}
\end{figure}

The average root mean square of the reconstruction error $\hat s - s$,
over all simulations for the Planck experiment, is computed in 10
bands of varying galactic latitude. The relative error, $r = E\left(s
- \hat{s}\right) / E\left(s\right)$, for both FastICA and the ILC, is
plotted in figures \ref{fig:rel_err_ica} and \ref{fig:rel_err_ilc}.

As we expected, FastICA is almost completely unaffected by calibration
errors.  Because no assumption on the relative calibration is used, the
overall calibration error just adds some small extra variance on the overall
level of the extracted CMB.

\begin{figure}
\includegraphics[width=85mm]{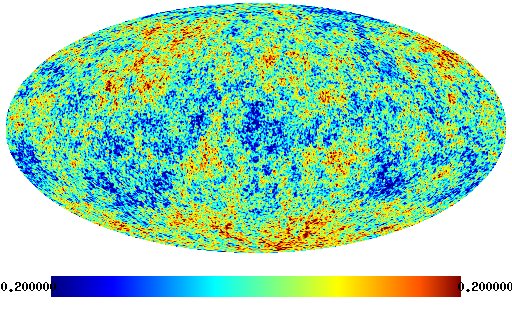}
\includegraphics[width=85mm]{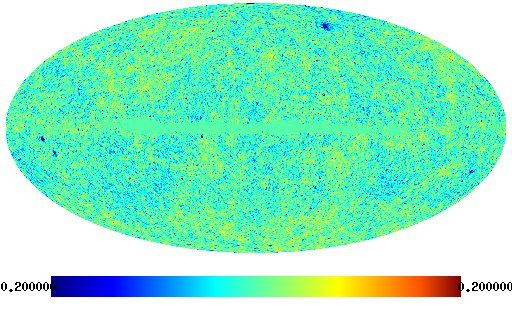}
\caption{Input CMB and ILC-estimated CMB plotted on a 0.2mK scale for one
  realization at 1\% calibration error with particularly bad output (relative
  error near 1.0).  Note that the variance of the ILC output is far below the
  input CMB, indicating that the input CMB was largely canceled.}
\label{fig:ilc_bad_example}
\end{figure}

ILC, however, is not so well behaved as FastICA.  While ILC is somewhat better
than FastICA at extracting the CMB when calibration is perfect, it quickly
becomes worse as calibration errors of increasing magnitude are applied.
Fig. \ref{fig:ilc_bad_example} shows the output of a particular realization at
1\% calibration error where ILC performed especially poorly, compared with the
input CMB plotted on the same scale.  The variance of the ILC output is much
lower than the true CMB, and CMB features are strongly suppressed.  As ILC
attempts to find the minimum-variance output, it finds that with calibration
errors it is possible to partially cancel the CMB to get the lowest possible
variance output.

\subsection{Interpretation of the ILC failure}

The impact of calibration errors on ILC weights, and on the output CMB map, is
analytically explored in Appendix \ref{app:ilcweights}. Here we highlight that
the signal-to-noise ratio plays a decisive role on this impact.

The ILC method is a linear combination of the maps observed in
different frequency channels, $\hat{s} = \sum_i w_i x_i.$ The ILC
combination has minimum variance under the constraint

\begin{equation}
\sum_i w_i a_i = 1.
\end{equation}

The constraint in principle guarantees the CMB conservation, otherwise
$w_i = 0$ for all $i$ would minimize the variance. If the calibration
$a_i$ is wrong then the CMB conservation is no longer guaranteed. In
some cases, when the signal-to-noise ratio is large enough, it can be
dramatic for the CMB extraction (see section \ref{subsec:sovern}).

\begin{figure}
\includegraphics[width=90mm]{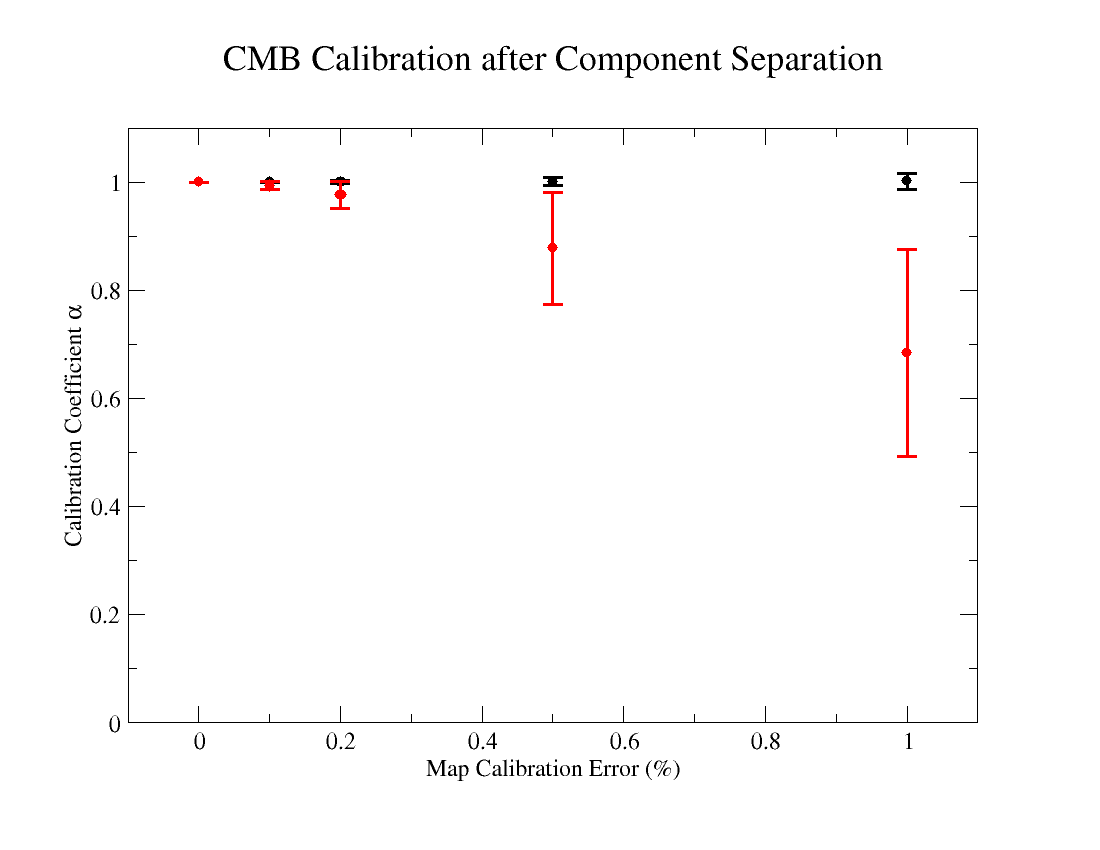}
\caption{This figure shows the overall calibration coefficient of the output
  CMB map, computed from the known calibration errors in the inputs and the
  weights applied to obtain the output.  The error bars represent the RMS of
  $\alpha$ among the 128 realizations.  For FastICA, the calibration
  coefficient is centered very near one, with an RMS of approximately 1.5
  times the map calibration error.  By contrast, ILC has a CMB calibration
  that is perfect if the map calibration is perfect, but this quickly turns
  into a significant bias with large uncertainties as to the final calibration
  value.}
\label{fig:cmb_contrib}
\end{figure}

As discussed above, the reconstruction error arises from two terms. A
multiplicative term, i.e. a global calibration error term, and an additive
contamination term.  We can write the estimated CMB map as a function of the
true CMB map as:
$$
\hat{s}(p)  = \alpha s(p) + c(p),
$$
where $\alpha$ is the global calibration coefficient, and $c(p)$ the
contamination by foregrounds and noise.  Figure
\ref{fig:cmb_contrib} shows this parameter $\alpha$ versus the input map
calibration error.
The presence of calibration errors $\delta_{a_i}$ modifies the calibration
coefficients in each channel as $a_i\rightarrow a_i+\delta_{a_i}$, where
$\delta_{a_i}\ll a_i$. We may explicitely expand the multiplicative error and
the additive error in terms of the calibration errors $\delta_{a_i}$ and the
ILC weights $w_i$:
\begin{eqnarray}
\hat{s}(p) & = & \sum_i w_i x_i(p)\cr
           & = & \sum_i w_i \left(a_i+\delta_{a_i}\right)s(p)+\sum_i w_i n_i(p)\cr
           & = & \left(1+\sum_i w_i\delta_{a_i}\right)s(p)+\sum_i w_i n_i(p),
\ea
where the ILC weights $w_i$ satisfy the constraint $\sum_i w_ia_i = 1$. Thus we have
\ba
c(p) & = & \sum_i w_i n_i(p),\\
\alpha & = & 1+\sum_i w_i\delta_{a_i}.
\end{eqnarray}

The additive error term $c(p)=\sum_i w_i n_i(p)$ is responsible for a bias in
the CMB estimation because of foreground and noise contaminations even in the
absence of calibration errors. Delabrouille et al. (2009) have explored the
impact of this term on the ILC estimation of the CMB and have found that, in
addition to the standard reconstruction error due to foreground and noise
contamination, there is a bias, $E\left(s\cdot\left(\hat{s}-s\right)\right)$,
due to the estimation of second order statistics on samples of finite
size. Both errors contribute to the variance of the output CMB map as
$$
E\left(\hat{s}^2\right) = E\left(s^2\right)+E\left(\left(\hat{s}-s\right)^2\right)+2E\left(s\cdot\left(\hat{s}-s\right)\right).
$$

The multiplicative error term $\alpha = 1+\sum_i w_i\delta_i$ becomes non
trivial in presence of calibration errors because the ILC weights $w_i$, as
derived in equation (\ref{qqq:weights}) of Appendix \ref{app:ilcweights}, do
not depend only on the calibration errors $\delta_{a_i}$ but also on the
signal-to-noise ratio $\sigma^2{R_n^{-1}}_{ij}$, where $\sigma^2 = E(s^2)$ and
$(R_n)_{ij}=E(n_i n_j)$ denote respectively the variance of the CMB signal and
the covariance matrix of the noise (including foregrounds).

\subsection{Importance of the signal to noise ratio}\label{subsec:sovern}
\label{sec:snr}

From the exact expression (\ref{qqq:weights}) of the weights we may write the multiplicative factor $\alpha  =  1+\bdw^t\bdelta_a$ as

\begin{eqnarray}\label{qqq:alpha}
\alpha & = & \frac{\bda^t\tR_n^{-1}\bda+\bda^t\tR_n^{-1}\bdelta_a}{\bda^t\tR_n^{-1}\bda+\sigma^2\left[(\bda^t\tR_n^{-1}\bda)(\bdelta_a^t\tR_n^{-1}\bdelta_a)-(\bda^t\tR_n^{-1}\bdelta_a)^2\right]}.\qquad
\end{eqnarray}
The immediate consequence of equation (\ref{qqq:alpha}) is the existence of
two regimes.

If the signal-to-noise ratio is small enough compared to the inverse of the
calibration error, typically,
$$
\mbox{if }\quad\sigma^2\bdelta_a^t\tR_n^{-1}\bdelta_a\ll 1\quad\mbox{then}\quad\alpha \approx 1+\mathcal{O}(\vert\bdelta_a\vert/\vert \bda\vert),
$$
because the expression proportional to $\sigma^2$ becomes negligible in
(\ref{qqq:alpha}). So we tend to recover the almost perfect CMB reconstruction
close to the case of no calibration error ($\bdelta_a = 0$).

If the signal-to-noise ratio becomes large enough then the reconstruction of
the CMB signal may be dramatically damaged. This is the main result of this
paper. Typically,
$$
\mbox{if }\quad\sigma^2\bdelta_a^t\tR_n^{-1}\bdelta_a\gg 1\quad\mbox{then}\quad\alpha \approx 0,
$$ the multiplicative factor goes to zero since the expression proportional to
$\sigma^2$ dominates all the other terms in (\ref{qqq:alpha}), in which case
the ILC estimation completely ``kills'' the expected CMB signal, $\hat{s}(p)
\approx c(p)$.

Let us complete the discussion by relating the first and second moments of the
output CMB $\hat{s}$ and the reconstruction error $\hat{s}-s$ to the
multiplicative and the additive errors. Considering that the CMB and the noise
(including foregrounds) are independent random signals, $E(n_is) = 0$, and
assuming that $E(n_i) = 0$, we get
\begin{eqnarray} \label{qqq:moments}
E(\hat{s}-s) & = &(\alpha-1)E(s),\cr
E\left(s\cdot\left(\hat{s}-s\right)\right) & = & (\alpha-1)E(s^2),\cr
E\left(\left(\hat{s}-s\right)^2 \right) & = & (\alpha-1)^2E(s^2)+E\left(c(p)^2\right),
\end{eqnarray}
where $E\left(c(p)^2\right) = \bdw^t R_n \bdw$. The detailed expression of
these moments in terms of the calibration errors and the signal-to-noise ratio
is derived in Appendix \ref{app:ilcweights}. From (\ref{qqq:moments}), once
again, if the signal-to-noise is large enough then the reconstruction of the
CMB is biased since $\alpha$ moves away from one to reach zero.

\subsection{A simple example}

Here we show a schematic
description of the process using a simple example.
We consider a two-channel case:

\begin{eqnarray}
x_1 &=& 0.99 s + n_1\cr
x_2 &=& s + n_2.
\end{eqnarray}

Here $s$ is the CMB, $x_i$ is the $i^{th}$ channel of the data, and $n_i$ is
the foregrounds plus instrument noise. The calibration coefficients are equal
to one and a calibration error of one percent has been considered in the first
channel.  If the signal-to-noise ratio is large enough, \emph{e.g} $n_i/s \ll
0.99$, then the noise is negligible in the observed maps
\begin{eqnarray}
x_1 &\approx& 0.99 s \cr
x_2 &\approx& s.
\end{eqnarray}
The ILC estimate of the CMB thus reduces in that case to
\begin{eqnarray}
\hat{s} & \approx & 100x_1-99 x_2,
\end{eqnarray}
where the weights satisfy the constraint $100-99=1$, which would guarantee the
CMB conservation if the calibration was correctly estimated. Consequently, the
CMB estimate is of minimum variance since $E(\hat{s}^2) \approx 0$, but of
course completely removes the expected input CMB, rendering the ILC totally
irrelevant.

 We may explain the process as follows. In presence of a calibration error in
 one channel the ILC algorithm minimizes the variance of
\begin{equation}
\hat{s} = (0.99 w_1 + w_2) s + w_1n_1 + w_2n_2.
\end{equation}
We can contrast this what we would get without calibration errors,
\begin{equation}
\hat{s} = (w_1 + w_2) s + w_1n_1 + w_2n_2.
\end{equation}
With the constraint that $w_1 + w_2 = 1$, the
contribution of the CMB signal to $\hat{s}$ is always $s$.  This
indicates that the weights will take whatever values they need to take
to minimize the contribution of the noise.

However, in the presence of calibration errors, it becomes possible
for the contribution of $s$ to $\hat{s}$ to vary depending upon the
choice of weights, indicating that a minimization of the variance of
$\hat{s}$ will introduce some competition between minimizing $(0.99
w_1 + w_2) s$ and minimizing $w_1n_1 + w_2n_2$.  For the following weights
\begin{eqnarray}
w_1 &=& 100,\cr
w_2 &=& -99,
\end{eqnarray}
the contribution of the CMB to $\hat{s}$ will be identically
zero.  This is what the ILC produces in the limit of the
signal-to-noise ratio becoming very large with respect to the
calibration error.  In the opposite limit, that of small
signal-to-noise ratio, it is the minimization of the second term,
$w_1n_1 + w_2n_2$, that drives the minimization of $\hat{s}$, which
mimics the behavior under the assumption of no calibration error.

\subsection{The case of Planck}

In table \ref{tab:planck} we present the results of ten simulations
of the sky with an ILC estimation of the CMB in presence of
$0.1~\%$, $0.5~\%$ and $1~\%$ calibration errors for the Planck
experiment ($9$ frequency channels).

\begin{table*}[htbp]
\begin{center}
\begin{tabular}{|p{6.5cm}|*{3}{c|}}
\hline
\emph{Planck}            & $1~\%$ & $0.5~\%$ & $0.1~\%$ \\
\hline \hline
\bfseries mult. factor $\boldsymbol{\alpha}$     & $0.66500$ & $0.85258$ & $0.99237$ \\
\hline
\bfseries  $\mbox{add. error }\boldsymbol{E(c(p))}\mbox{ (mK)}$  & $6.208$e$-2$ & $3.455$e$-2$ & $1.845$e$-2$\\
\hline
$E(c(p)^2)$ (${\mbox{mK}}^2$)   &  $1.231$e$-2$ & $3.46$e$-3$ &  $5.6$e$-4$\\
\hline \hline
$E((\hat{s}-s)^2)$  (${\mbox{mK}}^2$)               & $4.26$e$-3$ & $1.91$e$-3$ & $5.5$e$-4$\\
\hline
$E(s\cdot(\hat{s}-s))$ (${\mbox{mK}}^2$)           & $-3.26$e$-3$ & $-1.18$e$-3$ & $-1.3$e$-4$\\
\hline
$E(s^2)$  (${\mbox{mK}}^2$)                       & $7.42$e$-3$  & $7.42$e$-3$ & $7.42$e$-3$\\
\hline
$E(\hat{s}^2)$ (${\mbox{mK}}^2$)                 &   $5.16$e$-3$ & $6.99$e$-3$ &  $7.71$e$-3$\\
\hline
\end{tabular}
\end{center}
\caption{ILC reconstruction errors for Planck in presence of $1~\%$,
$0.5~\%$ and $0.1~\%$ calibration errors.} \label{tab:planck}
\end{table*}

For $1~\%$ we observe a significant bias affecting the CMB reconstruction by
ILC. The multiplicative factor $\alpha = 0.665$ (table \ref{tab:planck})
indicates that the CMB estimate eliminates roughly $33~\%$ of the input
CMB. The high sensitivity of Planck means a large signal-to-noise ratio,
comparable to the inverse of the calibration error, which leads to a poorly
extraction of the CMB by ILC, as expected from the formula (\ref{qqq:alpha}).
For $0.5~\%$ calibration errors, $15~\%$ of CMB is eliminated by the ILC
estimation. Finally for $0.1~\%$ calibration errors, $1~\%$ of CMB is
eliminated by the ILC estimation, which is nevertheless ten times the
calibration error -- and clearly not acceptable for precision cosmology with
Planck.

\subsection{The case of WMAP}

In table \ref{tab:wmap} we present the results of ten simulations of
the sky with an ILC estimation of the CMB in presence of $1~\%$
calibration errors for the WMAP experiment. We observe a negligible
bias affecting the CMB reconstruction by ILC. The multiplicative
factor $\alpha \approx 0.99$ (table \ref{tab:wmap}) indicates that the
percentage of eliminated input CMB by ILC is for WMAP of order of the
calibration error, \emph{i.e} $1~\%$, as expected from formula
(\ref{qqq:alpha}) when the signal-to-noise ratio is small enough.
The sensitivity of WMAP is small enough to render the ILC estimation
of the CMB insensitive to calibration errors.

\begin{table}[htbp]
\begin{center}
\begin{tabular}{|p{4.5cm}|*{3}{c|}}
\hline
\emph{WMAP}                       & $1~\%$ \\
\hline
\hline
\bfseries mult. factor $\boldsymbol{\alpha}$     &  $0.98709$    \\
\hline
\bfseries $\mbox{add. error }\boldsymbol{E(c(p))}\mbox{ (mK)}$     &  $1.129$e$-2$    \\
\hline
$E(c(p)^2)$ (${\mbox{mK}}^2$)            &  $6.9$e$-4$    \\
\hline
\hline
$E((\hat{s}-s)^2)$ (${\mbox{mK}}^2$)     &   $6.5$e$-4$   \\
\hline
$E(s\cdot(\hat{s}-s))$ (${\mbox{mK}}^2$) &  $-2.4$e$-4$    \\
\hline
$E(s^2)$ (${\mbox{mK}}^2$)               &  $5.15$e$-3$    \\
\hline
$E(\hat{s}^2)$ (${\mbox{mK}}^2$)         &  $5.33$e$-3$    \\
\hline
\end{tabular}
\end{center}
\caption{ILC reconstruction errors for WMAP in presence of $1~\%$
calibration errors.} \label{tab:wmap}
\end{table}

\subsection{Actual WMAP ILC}

The above result for WMAP was obtained assuming that the ILC is
performed on the masked sky of figure \ref{fig:mask}. In fact, ILC
weights used by the WMAP team have been computed in a different way, by
subdividing the sky into twelve regions. Since the value of their
weights are known, as well as the mean calibration error, we may easily
evaluate the error of the reconstruction performed by the WMAP team.

The order of magnitude of the ILC weights $w_i^{\rm WMAP}$ computed by the
WMAP team is comprised between $10^{-2}$ and $3$ \citep{hinshaw07} and the
relative calibration errors have been estimated by the WMAP team to be of the
order of $\delta_{a,i}\sim 0.2 \%$.

In the subdivision of the sky by the WMAP team the region zero
\citep{hinshaw07} corresponds to the part of the sky outside the galaxy and
thus dominated by the CMB signal. A priori, since the signal-to-noise ratio is
the highest in that high galactic latitude region, one might expect the effect
of the calibration errors to be large.  This is not the case, however.

We may estimate the maximum percentage of eliminated CMB in the region zero as
follows:
\begin{eqnarray}
\left\vert 1-\alpha \right\vert & = &\left\vert\sum_i w_i^{\rm WMAP}\delta_{a_i}\right\vert\cr
                     & \leq & 0.002\sum_i \left\vert w_i^{\rm WMAP}\right\vert \cr
                     & \leq & 7\cdot 10^{-3} = 0.7\%,
\end{eqnarray}
where $w_i^{\rm WMAP}$ are the ILC weights computed by the WMAP team in the
region zero \citep{hinshaw07}.  Therefore, the maximum percentage of
eliminated CMB has the order of magnitude of the calibration error, \emph{i.e}
${\rm few} \times 10^{-1}~\%$, which is small.\footnote{It should be noticed
  that this bound is a rough estimation since we do not have access to the
  real value of the calibration error for each frequency channel.} So the
multiplicative factor for the actual WMAP ILC in presence of $0.2~\%$
calibration errors is close to one, with a minor loss of CMB power:
$$
\alpha \geq 0.993.
$$

Interestingly, the ILC weights used at high galactic latitude by the WMAP team
\citep{hinshaw07} have been computed in a low galactic latitude region of the
sky, where the signal-to-noise ratio is sufficiently small. This certainly
explains why the ILC weights are close to those expected with no calibration
errors and why the multiplicative factor is close to one. Therefore the
calibration uncertainties do not have a strong impact on the ILC weights
computed in the WMAP third year data release. The bias due to calibration
errors is negligible.

The price paid for this, as emphasized by \citet{delabrouille09}, is that at
high galactic latitude the WMAP weights are chosen to cancel galactic
foregrounds rather than instrumental noise, a sub--optimal choice away from
the galaxy, particularly for small scales.

\subsection{Other ILC performed on WMAP}

Several authors have used a version of the ILC to analyze WMAP data. The
present paper warns the users of the corresponding data sets that in presence
of calibration errors, some CMB power may be lost in the maps
obtained. Further investigation would be needed to evaluate the exact impact
for each individual recovered CMB map.

\subsection{De--biasing}

A natural question to ask is whether, since the effect of calibration errors
is to introduce a loss of CMB power, it would not be possible to correct from
this effect and `recalibrate' a posteriori in some way.

First of all, this can not be the optimal solution, as the noise contribution
to the total error would be increased accordingly. The proper solution would
be to get the right calibration beforehand. As we can see from figure
\ref{fig:cmb_contrib} that the variance of $\alpha$ seems to be of order
$1-\alpha$.  This indicates that the maximum improvement on the level of the
CMB is to reduce the expectation value of $|1-\alpha|$ by around a factor of
two.  As $1 - \alpha$ becomes very large very quickly, this will not help when
the calibration is not already very good compared to the signal to noise ratio.

Finally, even the knowledge of the expectation value of $\alpha$ is not very
easy to get. Simulations give an estimate of its amplitude, but the actual
value may depend on details, for which simulations are not guaranteed to be
representative.

Hence, we leave this question open for further investigations.

\subsection{Impact of the number of channels}

Tests performed varying the number of channels used to perform the ILC with
Planck data show that ILC does better with calibration errors if fewer
channels are used.  To add new data, but end up with worse estimation of the
desired products, indicates that the new data is not being used effectively,
to say the least.

The reason for this degradation of the performance of the ILC when more
channels are added is easy to understand.  As discussed in section
\ref{sec:snr}, the ILC can erroneously cancel out part of the CMB if the
signal to noise ratio is larger than the inverse of the calibration error,
i.e. if
\begin{equation}
\sigma^2\bdelta_a^t\tR_n^{-1}\bdelta_a\gg 1.
\label{eq:snr-condition}
\end{equation}
As $\tR_n^{-1}$ and $\tR_n$ are symmetric matrices, they can be diagonalized,
and we can write:
$$\tR_n^{-1} = \tO^t \tD_n^{-1} \tO ,$$
where $\tO$ is an orthonormal matrix, and $\tD_n^{-1}$ a diagonal matrix. 
The condition of eq. (\ref{eq:snr-condition}) then becomes:
\begin{equation}
\sigma^2 (\tO \bdelta_a)^t \tD_n^{-1} (\tO \bdelta_a) \gg 1.
\end{equation}
Matrix $\tO$ preserves the norm, and thus elements of $\tO \bdelta_a$ are of
the same order as those of $\bdelta_a$.  It then suffices that one of the
eigenvalues of $\tD_n$ be small for $\sigma^2\bdelta_a^t\tR_n^{-1}\bdelta_a$
to be large, causing the CMB power loss discussed in this paper.

Now recalling that $\tR_n$ is the covariance matrix of noise + foregrounds, it
is easy to understand why more channels cause more problems with
Planck. Foregrounds are significantly brighter than the noise, and comparable
in amplitude to the CMB over a fraction of the sky. If they span a space of
dimension equal or greater than the number of channels, matrix $\tD_n$ will
have no small eigenvalue. If on the other hand they span a space of dimension
less than the number of channels, matrix $\tD_n$ will have at least one small
eigenvalue, generating the `CMB loss' problem.

Physically, this is understood in the following way: if there are few
channels, the minimization of the variance of the ILC linear combination will
be achieved by canceling foregrounds primarily. If however there are
additional channels which are not needed to cancel out the foregrounds, the
extra channels leave more freedom for the ILC weights to adjust themselves so
as to cancel part of the CMB as well.

\section{Conclusion}\label{sec:conclusion}

The primary conclusion of our work is that some care is required for
performing component separation in presence of calibration errors, in
particular for sensitive multichannel instruments such as Planck.  We have
shown that two different component separation algorithms, FastICA and ILC,
behave very differently in the presence of calibration errors.  FastICA is
completely unaffected, while ILC can become biased by a significant amount
with even small calibration errors.  We propose that those attempting to make
use of these or other component separation techniques pay close attention to
how calibration errors affect their results.  Some techniques will doubtlessly
be completely unaffected, as FastICA was, while others may be very sensitive
like ILC.

We also note that due to the fact that ILC in the presence of sufficient
calibration errors biases the variance of the CMB low, and because we have a
lower limit upon the variance of the CMB from WMAP, through its measurement of
the CMB power spectrum up to about $\ell=900$, the variance of the ILC output
may prove a useful diagnostic test if the calibration of Planck was performed
well.  The ability to use this as a cross-check on calibration also indicates
that for a Planck-style mission we expect to recover, at a minimum, around
0.1\%-0.2\% relative calibration error.  The reasoning for this is that if the
calibration error is worse, then ILC will produce a CMB map that is of
lower-variance than a similar map from WMAP, which, in turn, tells us that the
calibration wasn't very good.  If we have information that the calibration
wasn't as good as it could have been, then it is reasonable to expect that it
is possible to improve said calibration.

Note that even though FastICA is not biased where ILC is, it is not clear that
FastICA is better.  ILC does seem to produce lower errors in extracting the
CMB, as seen in figures \ref{fig:rel_err_ica} and \ref{fig:rel_err_ilc}.  The
biasing is troubling, but ILC retains lower extraction error up to somewhere
between 0.1\% and 0.2\% calibration error, at least at high galactic
latitudes.  If the calibration error is good enough, then we still expect
ILC to remain a very useful method for extracting the CMB.

\section*{Acknowledgments}
\thanks{
We would like to thank Carlo Baccigalupi, Jean-Fran\c{c}ois Cardoso,
Maude Le Jeune, and Radek Stompor for useful conversations related to
this work.}

\appendix
\onecolumn

\section{Analytic Analysis of ILC}
\label{app:ilcweights}

In this appendix we analytically derive the bias of the ILC estimator
generated by calibration errors, and look at the impact of the
signal-to-noise ratio on this bias. In particular we show that, even
in presence of small calibration errors, the ILC tends to poorly
extract the CMB if the signal-to-noise ratio is large.

\subsubsection*{Review of ILC}

We model the data as:
\begin{equation}
\bdx(p) = \bda ~s(p)+\bdn(p),
\end{equation}
where $\bdx(p)$ is the vector of the observed data at the pixel (or
harmonic mode, or needlet coefficient) $p$ for the set of frequency
channels, $s(p)$ is the CMB signal, and $\bdn(p)$ is the vector of
corresponding noise (including both foregrounds contaminants and
instrumental noise) for the set of frequency channels. $\bda$ is a
vector which contains the frequency scaling of the component, so that
in the case of CMB with no calibration errors one has $\bda =
(1,1,...,1)$. In the following we omit the index $p$.

We use $\sigma^2$ and $\tR_n$ respectively to denote the variance of the
CMB signal $s$ and the covariance matrix of the noise $\bdn$
(including foregrounds), and we assume that $s$ and $\bdn$ are
independent such that the covariance matrix of the observed maps is
$\tR_x=(\tR_n+\sigma^2\bda\bda^t)$.  The ILC implements an approximation
of the ideal filter
\begin{equation}
\hat{s} =\frac{\bda^t \tR_x^{-1} \bdx}{\bda^t \tR_x^{-1} \bda},
\end{equation}
which is an unbiased minimum variance estimate of $s$. In practice,
the covariance $\tR_x$ used in the ILC is an empirical estimate on a
sample of finite size, and thus slightly differs from its ensemble
average ($\hat{R}_x = \tR_x+\Delta_x$). This induces a bias in the
variance of the ILC, as shown by \citet{delabrouille09}. In this
appendix we assume $\hat{R}_x = \tR_x$ to investigate the bias that stems from errors in calibration alone ($\hat{\bda} = \bda+\bdelta_a$).

\subsubsection*{Bias due to calibration errors}

An imperfectly estimated frequency scaling vector $\bda$ includes
calibration errors $\bdelta_a$ which introduce a discrepancy between
the observed data and the ILC weights that are used to reconstruct the
CMB. This discrepancy is expected to be responsible for a bias in the
reconstruction.  Due to calibration errors $\bdelta_a$ the observed
map is modified as
\begin{eqnarray}
\bdx & = & (\bda+\bdelta_a) ~s+\bdn,\cr
\tR_x & = & \tR_n+\sigma^2(\bda+\bdelta_a)(\bda+\bdelta_a)^t,
\end{eqnarray}
such that the ILC estimate becomes\footnote{It should be noted that
  the derivation would have been of course completely equivalent if we
  had considered calibration errors into the ILC weights instead of
  into the data.}

\begin{equation}
\hat{s} =\frac{\bda^t
  \left[\tR_n+\sigma^2(\bda+\bdelta_a)(\bda+\bdelta_a)^t\right]^{-1}}{\bda^t
  \left[\tR_n+\sigma^2(\bda+\bdelta_a)(\bda+\bdelta_a)^t\right]^{-1}
  \bda}\left((\bda+\bdelta_a) ~s+\bdn\right).
\end{equation}

\noindent
Making use of the inversion formula:
\begin{equation}\label{qqq:inversion}
\tR_x^{-1}=\left[\tR_n+\sigma^2(\bda+\bdelta_a)(\bda+\bdelta_a)^t\right]^{-1}
=
\tR_n^{-1}-\sigma^2\frac{\tR_n^{-1}(\bda+\bdelta_a)(\bda+\bdelta_a)^t
  \tR_n^{-1}}{1+\sigma^2(\bda+\bdelta_a)^t
  \tR_n^{-1}(\bda+\bdelta_a)},
\end{equation}
we obtain the weights $\bdw$ of the ILC ($\hat{s}=\bdw^t\bdx$) as
\begin{equation}\label{qqq:weights}
\bdw^t =
\frac{\bda^t\tR_n^{-1}+\sigma^2\bda^t\tR_n^{-1}\left(\bda^t\tR_n^{-1}\bdelta_a+\bdelta_a^t\tR_n^{-1}\bdelta_a\right)-\sigma^2\bdelta_a^t\tR_n^{-1}\left(\bda^t\tR_n^{-1}\bda+\bda^t\tR_n^{-1}\bdelta_a\right)}{\bda^t\tR_n^{-1}\bda+\sigma^2(\bda^t\tR_n^{-1}\bda)(\bdelta_a^t\tR_n^{-1}\bdelta_a)-\sigma^2(\bda^t\tR_n^{-1}\bdelta_a)^2}.
\end{equation}

\noindent
In the absence of calibration errors ($\bdelta_a=0$) we recover the
standard ILC weights:
\begin{equation}
\bdw^t\vert_{\bdelta_a=0} = \frac{\bda^t\tR_n^{-1}}{\bda^t\tR_n^{-1}\bda} = \frac{\bda^t\tR_x^{-1}}{\bda^t\tR_x^{-1}\bda},
\end{equation}
where the second equality\footnote{In practice we have only access to
  the covariance matrix $\tR_x$ of the observed maps but not to the
  noise covariance matrix $\tR_n$ (including foregrounds) for
  constructing the ILC estimate. But theoretically the both
  representations of the ILC estimate are identical.} comes from the
inversion formula (\ref{qqq:inversion}).  We see that the presence of
calibration errors induces a departure from the standard ILC weights
through correction terms which explicitly depend on the
signal-to-noise ratio $\sigma^2\tR_n^{-1}$. Typically if the
signal-to-noise ratio is much smaller than the inverse of the
calibration error squared (\emph{e.g}
$\bdelta_a^t\sigma^2\tR_n^{-1}\bdelta_a\ll 1$) then the calibration
errors will have little impact and the standard ILC weights are
relevant. Else if the signal-to-noise ratio becomes comparable to the
inverse of the calibration error squared then the impact of
calibration errors may be more dramatic on the CMB reconstruction
since the variance of the ILC may be much lower than the true CMB. In
the exact expression (\ref{qqq:weights}) we intentionally conserved
second order terms in $\bdelta_a$ because they play a role of
regularization terms when the signal-to-noise ratio $\sigma^2\tR_n^{-1}$
goes to infinity.

As a simple illustration let us apply the above result to the
following example with two frequency channels and a diagonal noise
covariance matrix:

\begin{eqnarray}
x_1& = & 0.99 s+n_1\cr
x_2 &= & s+n_2.
\end{eqnarray}

Here the calibration error is one percent. We note $\sigma^2 =
E(s^2)$, $\tR_n = \mbox{diag}[\sigma_1^2,\sigma^2_2]$, where $\sigma_i^2
=E(n_i^2)$. In this example $\bda^t=(1,1)$ and $\bdelta_a^t=(-0.01,0)$
so that the expression (\ref{qqq:weights}) of $\bdw^t=(w_1,w_2)$
reduces to

\begin{eqnarray}
w_1 & = & \frac{\frac{\sigma_2^2}{\sigma^2}-\delta_{a1}}{\frac{\sigma_1^2+\sigma_2^2}{\sigma^2}+\delta_{a1}^2}\\
w_2 & = & \frac{\frac{\sigma_1^2}{\sigma^2}+(1+\delta_{a1})\delta_{a1}}{\frac{\sigma_1^2+\sigma_2^2}{\sigma^2}+\delta_{a1}^2}.
\end{eqnarray}

If the signal-to-noise ratio becomes very large (\emph{i.e}
$\sigma_i^2/\sigma^2\rightarrow 0$) then $w_1 \approx -1/\delta_{a1} =
100$ and $w_2 \approx (1+\delta_{a1})/\delta_{a1} = -99$, so that the
output CMB vanishes $\hat{s} \approx 100x_1-99x_2 \approx 100n_1-99n_2\approx 0$ when the noise is negligible. If the
signal-to-noise ratio becomes very small then
$w_1\approx\sigma_2^2/(\sigma_1^2+\sigma_2^2)$ and
$w_2\approx\sigma_1^2/(\sigma_1^2+\sigma_2^2)$, which is the standard
least mean square solution.

The ILC estimate is given by

\begin{eqnarray}
\hat{s}& =& \bdw^t(\bda+\bdelta_a)s +\bdw^t\bdn,\cr
       &  &\cr
\hat{s}& =& \frac{\bda^t\tR_n^{-1}\bda+\bda^t\tR_n^{-1}\bdelta_a}{\bda^t\tR_n^{-1}\bda+\sigma^2(\bda^t\tR_n^{-1}\bda)(\bdelta_a^t\tR_n^{-1}\bdelta_a)-\sigma^2(\bda^t\tR_n^{-1}\bdelta_a)^2}~s +\bdw^t\bdn,
\end{eqnarray}

We assume $E(\bdn)=0$, the ILC estimate is thus biased as

\begin{eqnarray}
E(\hat{s}) & =& \frac{\bda^t\tR_n^{-1}\bda+\bda^t\tR_n^{-1}\bdelta_a}{\bda^t\tR_n^{-1}\bda+\sigma^2(\bda^t\tR_n^{-1}\bda)(\bdelta_a^t\tR_n^{-1}\bdelta_a)-\sigma^2(\bda^t\tR_n^{-1}\bdelta_a)^2}~E(s).
\end{eqnarray}

We see that in the limit of small signal-to-noise ratio,
$\sigma^2\left(\tR_n^{-1}\right)_{ij}\ll 1$, the bias is of order of
magnitude of the calibration error:
$E(s)\bda^t\tR_n^{-1}\bdelta_a/\bda^t\tR_n^{-1}\bda$. Whereas if
$\sigma^2\left(\tR_n^{-1}\right)_{ij}\gg 1$ then the bias is accentuated
since $E(\hat{s})\rightarrow 0$.

Let us compute the mean value of the error $d = \hat{s}-s$ (assuming
$E(\bdn)=0$):
\begin{eqnarray}
E(d)& = & E(\bdw^t (\bda+\bdelta_a)s-s)\cr
    &   &\cr
    & = & \frac{\bda^t\tR_n^{-1}\bdelta_a-\sigma^2\left[(\bda^t\tR_n^{-1}\bda)(\bdelta_a^t\tR_n^{-1}\bdelta_a)-(\bda^t\tR_n^{-1}\bdelta_a)^2\right]}{\bda^t\tR_n^{-1}\bda+\sigma^2\left[(\bda^t\tR_n^{-1}\bda)(\bdelta_a^t\tR_n^{-1}\bdelta_a)-(\bda^t\tR_n^{-1}\bdelta_a)^2\right]}~E(s),
\end{eqnarray}
so that $E(d)\rightarrow -E(s)$ when
$\sigma^2\left(\tR_n^{-1}\right)_{ij}\gg 1$, and $E(d)\rightarrow 0$
when $\sigma^2\left(\tR_n^{-1}\right)_{ij}\ll 1$ and $\bdelta_a \ll
\bda$.

In the same way the cross correlation $E(s~d)$ of the error of the
reconstruction with the CMB signal
\begin{eqnarray}
E(s~d)& = & \frac{\bda^t\tR_n^{-1}\bdelta_a-\sigma^2\left[(\bda^t\tR_n^{-1}\bda)(\bdelta_a^t\tR_n^{-1}\bdelta_a)-(\bda^t\tR_n^{-1}\bdelta_a)^2\right]}{\bda^t\tR_n^{-1}\bda+\sigma^2\left[(\bda^t\tR_n^{-1}\bda)(\bdelta_a^t\tR_n^{-1}\bdelta_a)-(\bda^t\tR_n^{-1}\bdelta_a)^2\right]}~\sigma^2,
\end{eqnarray}
may vanish only if $\sigma^2\left(\tR_n^{-1}\right)_{ij}\ll 1$. If the
signal-to-noise ratio becomes large enough then $E(s~d)\rightarrow
-\sigma^2$, giving evidence of the cancellation of the CMB output.

We may also compute the variance of the error. Since the CMB and the noise are uncorrelated we get
\begin{eqnarray}
E(d^2)& = & \left(\bdw^t\bdelta_a\right)^2\sigma^2+\bdw^tR_n\bdw\cr
    &   &\cr
    & \approx & \left(\frac{\bda^t\tR_n^{-1}\bdelta_a-\sigma^2\left[(\bda^t\tR_n^{-1}\bda)(\bdelta_a^t\tR_n^{-1}\bdelta_a)-(\bda^t\tR_n^{-1}\bdelta_a)^2\right]}{\bda^t\tR_n^{-1}\bda+\sigma^2\left[(\bda^t\tR_n^{-1}\bda)(\bdelta_a^t\tR_n^{-1}\bdelta_a)-(\bda^t\tR_n^{-1}\bdelta_a)^2\right]}\right)^2~\sigma^2\cr
    &   &\cr
    &  & +\frac{\bda^t\tR_n^{-1}\bda+\sigma^2\left(2+\sigma^2\bda^t\tR_n^{-1}\bda\right)\left[(\bda^t\tR_n^{-1}\bda)(\bdelta_a^t\tR_n^{-1}\bdelta_a)-(\bda^t\tR_n^{-1}\bdelta_a)^2\right]}{\left(\bda^t\tR_n^{-1}\bda+\sigma^2\left[(\bda^t\tR_n^{-1}\bda)(\bdelta_a^t\tR_n^{-1}\bdelta_a)-(\bda^t\tR_n^{-1}\bdelta_a)^2\right]\right)^2}.
\end{eqnarray}
Notice that the second term has been truncated at second order in
$\bdelta_a$. If $\sigma^2\left(\tR_n^{-1}\right)_{ij}\ll 1$ then we recover
the standard reconstruction error with no calibration error: $E(d^2) \approx
1/\left(\bda^t\tR_n^{-1}\bda\right)$, as computed by
\citet{delabrouille09}. If the signal-to-noise ratio becomes large enough then
$\bdw^tR_n\bdw$ becomes negligible compared to the first term
$\left(\bdw^t\bdelta_a\right)^2\sigma^2$ such that $E(d^2)\approx \sigma^2$,
again giving evidence of the cancellation of the CMB output.

\bibliography{./cmb}

\label{lastpage}

\end{document}